\documentclass{fizik} 
\vol{} \pyear{} \received{}
\newcommand{\ben}{\begin{enumerate}}
\newcommand{\een}{\end{enumerate}}
\newcommand{\be}{\begin{equation}} \newcommand{\ee}{\end{equation}}
\newcommand{\bse}{\begin{subequation}}\newcommand{\ese}{\end{subequation}}
\newcommand{\bea}{\begin{eqnarray}} \newcommand{\eea}{\end{eqnarray}}
\newcommand{\bc}{\begin{center}} 
\newcommand{\ec}{\end{center}}

\def\erf{\mathop{\rm erf}\nolimits}
\def\ln{\mathop{\rm ln}\nolimits}

\author       
{        {\bf  Maia ANGELOVA }\\        {\it School of Computing and Mathematics}\\         {\it  University of Northumbria, Newcastle upon Tyne, UK GB-NE1 8ST}\\        
{\bf A. FRANK}\\         
{\it Instituto de Ciencias Nucleares and Centro de Ciencias F\'{i}sicas,}\\      
{\it  UNAM, A.P. 70-543, Mexico, D.F., 04510 Mexico.}\\    }

\title{Anharmonic Algebraic Model of Thermodynamic Properties of Diatomic Molecules}

\begin{document}
\maketitle

\begin{abstract}

An algebraic model based on Lie-algebraic techniques is applied to the analysis of thermodynamic vibrational properties of diatomic molecules. The local anharmonic effects are described by a Morse-like potential and corresponding  anharmonic bosons associated with the $U(2)$\ algebra. A vibrational high temperature  partition function and the related thermodynamic functions  are derived in terms of the parameters of the model. The thermal expansion operator is defined and   obtained using the renormalized frequency of the model. 

\end{abstract}

\section{Anharmonic Algebraic Model}

Algebraic models have been used very successfully in nuclear physics and
have led to new insights into the nature of complex many body systems [1].
Similar methods have been proposed to describe molecular excitations [2,3].
The methods combine Lie algebraic techniques, describing the interatomic
interactions, with discrete symmetry techniques associated with the molecules.
In the framework of the algebraic model [2], the anharmonic effects of the
local interactions are described by substituting the local harmonic potentials by
Morse-like potential, associated with the $U(2)$ algebra. The
one-dimensional Morse Hamiltonian is written in terms of the \bigskip
generators $\ $of $U(2)$,

\begin{equation}
H_{M}=\frac{A}{4}\left( \hat{N}^{2}-4\hat{J}_{Z}^{2}\right) =\frac{A}{2}(%
\hat{J}_{+}\hat{J}_{-}+\hat{J}_{-}\hat{J}_{+}-\hat{N})
\end{equation}
where $A$ is a constant. The eigenstates, $\mid \![N],v\rangle $, correspond to the  
$\ U(2)\supset SU(2)$ symmetry-adapted basis,$\;$where $%
N\; $is the total number of bosons fixed by the potential shape, and $v$ is
the number of quanta in the oscillator, $v=1,2,\ldots ,\left[\frac{N}{2}\right]$. 

The anharmonic effects are described by anharmonic boson
operators [2],

\begin{equation}
\hat{b}=\frac{\hat{J}_{+}}{\sqrt{N}},\;\;\;\hat{b}^{\dagger }=\frac{\hat{J}%
_{-}}{\sqrt{N}}\;,\;\;\;\hat{v}=\frac{\hat{N}}{2}-\hat{J}_{z}
\end{equation}
where $\hat{v}\;\;$is the Morse phonon operator with an eigenvalue $v$. The
operators satisfy the commutation relations,

\begin{equation}
\left[ \hat{b},\hat{v}\right] =\hat{b},\;\;\;\;\;\;\;\left[ \hat{b}^{^{\dagger }},%
\hat{v}\right] =-\hat{b}^{^{\dagger }},\;\;\;\;\left[ \hat{b},\hat{b}^{^{\dagger }}%
\right] =1-\frac{2\hat{v}}{N}
\end{equation}

The harmonic limit is obtained when $\;\;N\rightarrow \infty $, \ in which
case \ \ $\left[ \hat{b},\hat{b}^{^{\dagger }}\right] \rightarrow 1$ \ giving the usual
boson commutation relations.

The Morse Hamiltonian can be written in terms of \ the operators $\hat{b}$ and 
$\hat{b}^{^{\dagger }}$,

\begin{equation}
H_{M}\sim \frac{1}{2}\left( \hat{b}\hat{b}^{^{\dagger }}+\hat{b}^{^{\dagger
}}\hat{b}\right)
\end{equation}
which corresponds to vibrational energies

\begin{equation}
\varepsilon _{v}=\hbar \omega _{0}\left( v+\frac{1}{2}-\frac{v^{2}}{N}%
\right) .
\end{equation}
where $\omega _{0}$ is the harmonic oscillator frequency. 
The Morse potential spectrum thus  leads to a deformation of the  harmonic oscillator algebra. A more detailed relationship between the Morse coordinates and momenta and the $SU(2)$ generators can be derived through a comparison of their matrix elements [5]. 
Note that for an infinite potential depth, $N\rightarrow \infty $, the Morse potential
cannot be distinguished from the harmonic potential.

The anharmonic model has been developed to analyze  molecular vibrational
spectra [2-8]. It provides a systematic procedure for studying
vibrational excitations in a simple form by describing the stretching and
bending modes in a unified scheme based on $U(2)$ algebras which incorporate
the anharmonicity at the local level.

The aim of this paper is to apply the algebraic approach to the vibrational 
high-temperature 
thermodynamics of a diatomic molecule and obtain the basic thermodynamic
functions in terms of the parameters of the anharmonic model.
These results must be combined with the translational and rotational thermodynamic functions in order to compare with experiment, in a fashion similar to that of reference [9]. Here, we concentrate our attention on the derivation of analytic results which will be applied in a subsequent paper [10]. 

\section{Vibrational Partition Function}

The vibrational partition function  of a diatomic anharmonic molecule is

\begin{equation}
Z_{N}=\sum_{v=0}^{[N/2]}e^{-\beta \varepsilon _{v}}
\end{equation}
where $\beta =1/k_{B}T$, the vibrational energies $\varepsilon _{v}$ are
given in  equation (5) and $N$ is the fixed total number of anharmonic
bosons.

Introducing new parameters,$\;\alpha =\frac{\beta \hbar \omega _{0}}{2}%
,\;N_{0}=\left[\frac{N}{2}\right]$ and $\;l=\left[\frac{N}{2}\right]-v$, the vibrational partition
function can be written as,

\begin{equation}
Z_{N}=e^{-\alpha }\sum_{l=0}^{N_{0}}e^{-\frac{\alpha }{N_{0}}\left(
N_{0}^{2}-l^{2}\right) }.
\end{equation}

At high temperatures $T$, for  $N_{0}$  large and  $\alpha $  small, the sum can be replaced by the integral,

\begin{equation}
Z_{N}=\sqrt{\frac{N_{0}}{\alpha }}e^{-\alpha \left( N_{0}+1\right)
}\int\limits_{0}^{\sqrt{\alpha N_{0}}}e^{s^{2}}ds
\end{equation}
where $s=\sqrt{\frac{\alpha }{N_{0}}}l$. This integral can be evaluated exactly in terms of 
the error function,  
$\erf{i\!\left( \sqrt{\alpha N_{0}}\right)}$ (see [11]), 

\begin{equation}
Z_{N}=\frac{1}{2}\sqrt{\frac{N_{0}{\pi }}{\alpha }}
e^{-\alpha \left(N_{0}+1\right) }{\erf i\!\left( \sqrt{\alpha N_{0}}\right)} .
\end{equation}

Equation (9)  represents the high-temperature value of the vibrational partition
function in the Morse-like spectrum. 
When $N_{0}\rightarrow \infty,\;\;\alpha N_{0}\gg 1$, the harmonic limit of the model is obtained,

\begin{equation}
Z_{\infty }\sim \frac{N_{0}e^{-\alpha }}{2\alpha N_{0}-1}\sim \frac{%
e^{-\alpha }}{2\alpha }=\frac{k_{B}T}{\hbar \omega _{0}}e^{-\frac{\hbar
\omega _{0}}{2k_{B}T}}
\end{equation}
which coincides with the harmonic vibrational partition function 
of a diatomic molecule at high temperatures.

Having the partition function $Z_{N}$ \ in terms of the parameters of the algebraic model, we are now in position to derive the basic thermodynamic functions. An algebraic approach has been used in [12] to study the thermodynamic properties of molecules. However, the partition function in [12] uses an approximation of the  classical density of states, while we have derived an explicit function in terms  of the parameters of the algebraic model which can now be used to compute the high-temperature thermodynamic functions in analytic form. 

\section{Thermodynamic Vibrational Functions}

\subsection{Mean Vibrational Energy}

The mean vibrational energy is given by

\begin{equation}
U_{N}=-\frac{\partial }{\partial \beta }\ln Z_{N}=-\frac{\hbar \omega _{0}}{%
2Z_{N}}\frac{\partial Z_{N\;}}{\partial \alpha }.
\end{equation}

Taking into account that 
\begin{equation}
\frac{\partial Z_{N\;}}{\partial \alpha }=-\frac{Z_{N}}{2\alpha }-\left(
N_{0}+1\right) Z_{N}+\frac{N_{0}e^{-\alpha }}{2\alpha }
\end{equation}
we obtain the following expression for the mean vibrational energy in terms
of the partition function $Z_{N}$,

\begin{equation}
U_{N}=\frac{\hbar \omega _{0}}{2}\left( 1+N_{0}+\frac{1}{2\alpha }-\frac{%
N_{0}e^{-\alpha }}{2\alpha Z_{N}}\right) .
\end{equation}
Substituting  $Z_{N}$ by expression (9) gives,
\begin{equation}
U_{N}=\frac{\hbar \omega _{0}}{2}(1+N_{0})+\frac{k_{B}T}{2}\left( 1-\sqrt{%
\frac{\hbar \omega _{0}N_{0}}{2\pi k_{B}T}}\frac{e^{-\frac{N_{0}\hbar \omega
_{0}}{2k_{B}T}}}{\erf i\!\left( \sqrt{\frac{N_{0}\hbar \omega _{0}}{%
2k_{B}T}}\right) }\right) .
\end{equation}

The harmonic limit is obtained from equation (13), when 
$N_{0}$ $\rightarrow \infty $ and $Z_{N}$ is given by (10),

\begin{equation}
U_{\infty }\sim \frac{\hbar \omega _{0}}{2}+k_{B}T.
\end{equation}
This is the classical mean energy of a diatomic molecule at very high
temperatures.

\subsection{Specific Heat}

The vibrational part of the specific heat is,

\begin{equation}
C_{N}=\frac{\partial U_{N}}{\partial T}=-\frac{\hbar \omega _{0}}{2k_{B}T^{2}%
}\frac{\partial U_{N}}{\partial \alpha }.
\end{equation}
Substituting $U_{N}$ with equation (13) and using (12), we obtain

\begin{equation}
C_{N}=\frac{k_{B}}{2}+\frac{k_{B}N_{0}e^{-\alpha }}{2Z_{N}}\left(
N_{0}\alpha -\frac{1}{2}-\frac{N_{0}e^{-\alpha }}{2Z_{N}}\right)
\end{equation}
This represents the vibrational specific heat in the anharmonic model in
terms of the partition function $Z_{N}$. All anharmonic contributions depend
on the temperature. The explicit value is obtained from equation
(17) by substituting $Z_{N}$ by expression (9).  
When $N_{0}$ $\rightarrow \infty $, the harmonic limit of the model  gives the vibrational specific heat of a diatomic molecule at very high temperatures,   

\begin{equation}
C_{\infty }\sim k_{B}.
\end{equation}

\subsection{Mean Number of Anharmonic Bosons}

The mean vibrational energy in the anharmonic model can be written in terms
of mean number\ $\langle \nu _{N}\rangle $ of anharmonic quanta, each with
energy $\hbar \omega_{0} $,

\begin{equation}
U_{N}=\hbar \omega _{0}\left( \langle \nu _{N}\rangle +\frac{1}{2}\right)
\end{equation}
Substituting $U_{N}$ by equation (13), we obtain $%
\langle \nu _{N}\rangle $ in terms of the partition function $Z_{N},$

\begin{equation}
\langle \nu _{N}\rangle =\frac{N_{0}}{2}+\frac{1}{4\alpha }-\frac{%
N_{0}e^{-\alpha }}{4\alpha Z_{N}}.
\end{equation}

Using expression (9) in equation (20), we obtain the high-temperature value, 

\begin{equation}
\langle \nu _{N}\rangle =\frac{N_{0}}{2}+\frac{k_{B}T}{2\hbar \omega }-\sqrt{%
\frac{N_{0}}{\pi }}\left( \frac{k_{B}T}{\hbar \omega }\right) ^{\frac{1}{2}}%
\frac{e^{N_{0\frac{\hbar \omega }{2k_{B}T}}}}{\erf i\!\left( \sqrt{\frac{%
\hbar \omega N_{0}}{2k_{B}T}}\right) }.
\end{equation}

The harmonic limit is obtained from equation (20) when $N_{0}\rightarrow \infty$ and $Z_{N}$ is given by expression (10),

\begin{equation}
\langle \nu _{\infty }\rangle \sim \frac{k_{B}T}{\hbar \omega _{0}}.
\end{equation}

\subsection{Free Energy}

The free vibrational energy in terms of the partition function $Z_{N}$ is
given by

\begin{equation}
F_{N}=-\frac{1}{\beta }\ln Z_{N}
\end{equation}

Substituting $Z_{N}$ with equation (9) gives the free vibrational energy in
the algebraic model at high temperatures,

\begin{equation}
F_{N}=k_{B}T\ln 2+k_{B}T\ln \left({\frac{\hbar \omega _{0}}{2\pi N_{0}k_{B}T}}\right)-%
\frac{\hbar \omega _{0}}{2}\left( N_{0}+1\right) -k_{B}T\ln \left( \erf%
i\!\sqrt{\frac{\hbar \omega _{0}N_{0}}{2k_{B}T}}\right) .
\end{equation}

Using  expression (10) in equation (23), we obtain the classical harmonic result for the free vibrational energy at very high temperatures,
\begin{equation}
F_{\infty }\sim k_{B}T\ln 2 .
\end{equation}

\section{Thermal Expansion}

It is well-known that the thermal expansion in a molecule is a result of the
local anharmonic effects and cannot be explained by the harmonic models. It depends on the positions of the atoms and their displacement from the equilibrium. As an approximation, 
we can write the coordinate operator $\hat{x}$ in terms of the anharmonic operators $\hat{b}$ and $\hat{b}^{^{\dagger }}$  (see also [5]) as 

\begin{equation}
\hat{x}=\sqrt{\frac{\hbar }{2\mu \omega }}\left( \hat{b}+\hat{b}^{\dagger }\right)
\end{equation}
Here, $\omega $ depends on the number of quanta in the
oscillator [4-7],

\begin{equation}
\omega =\omega _{0}\left( \frac{N+1}{N}-\frac{2\nu }{N}\right)
\end{equation}
\qquad

Now, we define the thermal expansion operator, 

\begin{equation}
\hat{X}_{N}=\sqrt{\frac{\hbar }{2\mu \omega }}\left( \hat{b}+\hat{b}^{\dagger }+\frac{\gamma 
}{N}\right) 
\end{equation}
where $\frac{\gamma }{N}$ \ represents the displacement from the equilibrium
position. The harmonic limit is obtained  when $N\rightarrow \infty $, in which case  
$\omega \rightarrow \omega _{0}$ and $\;\frac{\gamma }{N}\rightarrow 0$.

Evaluating the mean value of the operator, $\hat{X}_{N}$, between the states $\mid\![N]v\rangle $ and taking into account (27), we obtain the mean
thermal expansion $X_{N}$ in terms of the parameters of the anharmonic model,

\begin{equation}
X_{N}=\sqrt{\frac{\hbar }{2\mu \omega _{0}}}\frac{\gamma }{\sqrt{N\left(
N+1\right) }}\left( 1-\frac{2\upsilon }{N+1}\right) ^{-\frac{1}{2}}
\end{equation}

The anharmonic effects are essential at high temperatures $T$. When 
$N{\gg}v$, equation (29)  becomes,

\bigskip 
\begin{equation}
X_{N}\sim \sqrt{\frac{\hbar }{2\mu \omega _{0}}}\frac{\gamma }{N}\left( 1+\frac{v%
}{N}\right) 
\end{equation}
Taking into account the high-temperature value of the mean number of quanta (22), 

\begin{equation}
X_{N}\sim \sqrt{\frac{\hbar }{2\mu \omega _{0}}}\frac{\gamma }{N}\left( 1+\frac{%
k_{B}T}{\hbar \omega _{0}N}\right) .
\end{equation}

The thermal expansion of anharmonic diatomic molecule has been obtained in [13], using a Morse-type potential and classical methods. The comparison of  equations (29) and (30)-(31) with the results in [13] at high temperatures shows very good agreement. It also shows that $ \frac{\gamma }{N}$ is proportional to the mean separation between the atoms at $T=0$ and that $N^{2}$ is proportional to the depth of the potential well, the latter being in agreement with the algebraic model [4-7]. 

\section{Conclusion} We have applied the algebraic model to those thermodynamic 
properties of diatomic molecule which at high temperature  strongly depend on its anharmonicity. We have derived the vibrational partition function and the related thermodynamic functions, such as mean vibrational energy and specific heat, in terms of the parameters of the model. The next step is to  consider the rotational excitations  
of the molecule. The thermal expansion, which  is a typical anharmonic effect, has been  derived  in terms of the parameters of the model and its high-temperature value is  in a good agreement with classical anharmonic results. The model can be  further  applied to the thermodynamics of polyatomic molecules.  
This and other applications will be discussed elsewhere. 

\section*{Acknowledgements} This work was supported in part by project (32397-E) Conacyt, Mexico. 

\begin{reference}
\item F. Iachello and A. Arima, {\it The Interacting Boson Model } (CUP
1987).
\item  A. Frank A and P. Van Isacker,{\it Algebraic Methods in Molecular
and Nuclear Structure Physics } (John Wiley 1994) .
\item  F. Iachello and R. Levine, {\it Algebraic Theory of Molecules } (OUP
1995).
\item  R. Lemus and A. Frank, {\it J. Chem. Phys.} {\bf 101} (1994) 8321.
\item A. Frank, R. Lemus, M. Carvajal, C. Jung and E. Ziemniak, {\it Chem. Phys. Lett. \bf 308} (1999) 91.  
\item A. Frank, R. Lemus, R. Bijker, F. P\'{e}rez-Bernal and J.M. Arias,  {\it Annals of Physics} {\bf 252} (1996) 211.
\item F. P\'{e}rez-Bernal and J.M. Arias, A. Frank, R. Lemus and R. Bijker, {\it J. Mol. Spectrosc.} {\bf 184} (1997) 1.
\item S. Oss, {\it Adv. Chem. Phys. \bf 43} (1996) 455. 
\item G. Herzberg, {\it Molecular Spectra and Molecular Structure}, volume II 
(Van Nostrand 1945). 
\item M. Angelova and A. Frank, to be published. 
\item  {\it Handbook of Mathematical Functions}, Edited by M. Abramowitz and I.A. Stegun (National Bureau of Standards 1964). 
\item  D. Kusnetzov, {\it J. Chem. Phys., }\textbf{101 }(1994) 2289.
\item P. Br\"{u}esch. \textit{Phonons: Theory and Experiments}, volume 
\textbf{1} (Springer 1992).

\end{reference}

\end{document}